\documentclass{iau}
\usepackage{graphicx}

\title[Future of Pulsar Research and Facilities] %% give here short title %%
{The Future of Pulsar Research and Facilities}

\author[Matthew Bailes]   %% give here short author list %%
{Matthew Bailes$^{1,2}$
}

\affiliation{$^1$Centre for Astrophysics and Supercomputing, Swinburne University of Technology, \\ H11 PO Box 218,
Hawthorn, Australia, 3122 \\ email: {\tt mbailes@swin.edu.au} \\[\affilskip]
$^2$ARC Centre of Excellence for Gravitational Wave
Discovery (OzGrav)}

\pubyear{2017}
\volume{337}  %% insert here IAU Symposium No.
\pagerange{XXX--XXX}
\jname{Title of your IAU Symposium}
\editors{A.C. Editor, B.D. Editor \& C.E. Editor, eds.}
\begin{document}

\maketitle

\begin{abstract}
Radio pulsars have been responsible for many astonishing astrophysical and fundamental physics breakthroughs since their discovery 50 years ago. In this review I will discuss many of the highlights, most of which were only possible because of the provision of large-scale observing facilities. The next 50 years of pulsar astronomy can be very bright, but only if our governments properly plan and fund the infrastructure necessary to enable future discoveries. Being a small sub-field of astronomy places an onus on the pulsar community to have an open-source/open access approach to data, software, and major observing facilities to enable new groups to emerge to keep the field vibrant.

\keywords{pulsars,  radio pulsars, radio telescopes.}
\end{abstract}

\firstsection % if your document starts with a section,
\section{Introduction}

Michael Kramer once wrote in one of our funding applications that, ``There is arguably no other sub-domain of astronomy that has such a broad  impact across so many areas of fundamental physics as pulsar astronomy''. It was a good line, the proposal was funded, and I've been cutting and pasting it ever since. The reason it is such a
great line is because it is probably true. Pulsar astronomy lies at the intersection between astronomy and fundamental physics. This is its great strength and unlocks considerable resources.

Pulsars have another attribute that makes working in the field an absolute joy, and that is that most of us really love them! When I was an impressionable graduate student I remember visiting the then proto-Nobel Laureate Joe Taylor. He explained it to me thus, ``My work is my hobby''. Many of us in the field can relate to Joe's statement. Because pulsars are distinct objects, there is a special joy when a search pipeline uncovers a new one, and another when the confirmation observation is positive. Is it a binary? Is it young? Will this pulsar be special??? There are also few greater joys in life than when one first obtains a phase-coherent timing solution across a long series of hard-won observations! Pulsar timing is at the heart of most of the field's greatest achievements, and has led to many incredible breakthroughs since 1967.

Pulsars were first discovered 50 years ago, and in this symposium we had the privilege of hearing from Jocelyn Bell about how she and Antony Hewish developed the instrument that made this possible, and her detective work that led to the discovery.  It was the only talk I have ever heard at a conference that led to a standing ovation - even though we were in Britain! I'm not sure if this talk was recorded, but it should be compulsory viewing for anyone entering the field. 

Many of us are upset that Jocelyn didn't share in the Nobel prize that was ultimately awarded for this discovery because we recognise that it was more than just hard work that enabled it. It required many late and cold nights, an eye and memory for detail, some degree of stubbornness in the face of skepticism, and the kind of expert knowledge of an instrument that is usually only achieved by those who have built and debugged it. Pulsar astronomy has only advanced because of the constant renewal of hardware (and now software) and many hours using our telescopes. Anyone who has ever commissioned or developed a new pulsar instrument can relate to Jocelyn's story. Although she may not have a Nobel prize she certainly has the respect and admiration of the entire pulsar community.

In this paper I will first review some of the first 50 years of pulsar astronomy's greatest contributions to science, before pondering if anything is left to be discovered and how I think it will evolve in the future. Finally, I will explore the interplay between science and recent political developments in the era of the Internet, and how difficult it is to plan and execute breakthrough science that involves 100M+ infrastructure in the era of fake news and populism.

\section{Great Pulsar Discoveries}

The greatest pulsar discovery of all was of course the discovery of the first four pulsars themselves, described in \cite[Hewish et al. (1968)]{Hew1968} that at the time of this lecture has 698 citations. The 1968 Nature paper triggered a wave of pulsar surveys around the world that soon revealed that pulsars were a galactic population, and it wasn't long before  the youngest pulsars were associated with the Crab \cite[(Staelin \& Reifenstein 1968)]{staelin1968} and Vela \cite[(Large et al. 1968)]{large1968} supernova remnants. The Crab's 33 ms period eliminated many non-neutron star pulsar models and it was soon established beyond doubt that pulsars were neutron stars. The brevity of the abstract of the Large et al. Nature paper is an indication of its significance and perhaps a lost art, \\
\\
{\centering{\it{``A pulsar with a very short period (0.089 s) has been discovered at the position of a suspected supernova remnant, raising several interesting consequences.''}}}
\\
\\
a lesson for all of us, (especially Jim!).

It wasn't long before the first glitch of a pulsar was observed in the Vela pulsar by \cite[Radhakrishnan \& Manchester (1968)]{rad1968} (see Manchester these proceedings) and in the following year \cite[Radhakrishnan \& Cooke (1969)]{rad1969} introduced the rotating vector model (RVM) that described the observed polarisation angle swing of pulsars in terms of misaligned magnetic field and spin axes, and the line of sight to Earth. Glitches provide an avenue with which to explore the superfluid interiors of pulsars, whilst the RVM model gives some indication of the orientation of pulsar spin and magnetic field axes.

Although more than half of all OB (massive) stars are members of binary systems it was a great mystery as to why the first 70-odd pulsars were solitary? The first pulsars were found using chart recorders, but Joe Taylor pioneered the use of the fast fourier transform combined with an ingenious dedispersion technique (the tree algorithm) \cite[(Taylor 1974)]{taylor1974} and the might of the Arecibo 305\,m dish to explore new phase space \cite[(Hulse \& Taylor 1974)]{hulse1974}. This enabled coherent integrations that lowered the effective minimum flux a pulsar could possess and still be detectable. The jewel in this survey was PSR B1913+16, the first binary pulsar discovered \cite[(Hulse \& Taylor 1975)]{hulse1975}. PSR B1913+16 possessed a 7.75h orbit, an orbital eccentricity of 0.617 and ultimately led to convincing proof that gravitational waves existed \cite[(Weisberg \& Taylor 1984)]{weisberg1984} (see Taylor these proceedings). It was the initial motivation to build LIGO, and
the impetus for the study of massive binary evolution, pulsar spin-up (see Bhattacharya these proceedings) and has proved to be an exhaustive test-bed for theories of relativistic gravity. It ultimately won the Nobel prize for its discoverers, Hulse and Taylor.

The discovery of the 1.56 millisecond pulsar (MSP) \cite[(Backer et al. 1982)]{backer1982} was another watershed moment for the field. Until this time, the fastest pulsar known was the Crab pulsar, with a spin period of 33 ms. The millisecond pulsar was important for two reasons. First, its inferred magnetic field strength was four orders of magnitude lower than the ``normal'' pulsar population, and it opened the possibility that pulsar timing could be pushed to sub-microsecond precision.

\cite[Lyne, Anderson \& Salter (1982)]{las82} performed the
first comprehensive study of pulsar proper motions using
an interferometer. They demonstrated that pulsars were
high velocity objects with speeds of up to
$\sim 300 $km s$^{-1}$, were leaving the galactic plane,
and that the supernovae that created them were possibly
asymmetric. Subsequently \cite[Cordes, Romani \& Lundgren (1993)]{cordes1993}
found evidence that the pulsar in the so-called ``guitar nebula'' had a velocity of order 1000 km s$^{-1}$. \cite[Lyne \& Lorimer (1994)]{lyne1994}
used some simple arguments to show that young pulsars had velocities
of up to 500 km s$^{-1}$.

\cite[Kulkarni (1986)]{kulkarni1986} was the first to use optical telescopes to study the white dwarf companions of binary pulsars to dramatic effect. Up until this time it was widely believed that pulsar
magnetic fields decayed exponentially with a timescale of $\sim$5 Myr.
The very cool white dwarf companion of PSR B0655+64 demonstrated
that this was false. If its field was decaying the white dwarf should
have been very young. Evidently, field decay in the spun-up
pulsars didn't appear to be active. This was part of a body of evidence that led some of us to later argue e.g.  \cite[(Bailes 1989)]{bailes1989} that
accretion is the dominant source of field decay in pulsars, and that
isolated pulsars exhibit little or none. 

On evolutionary grounds it was argued that millisecond pulsars were the descendants of low-mass X-ray binaries \cite[(Alpar et al. 1982)]{alpar1982} and this made globular clusters tempting targets for pulsar surveys.
The historic first detection was made by \cite[Lyne et al. (1987)]{lyne1987}, who found a 3\,ms pulsar in M28. Subsequent surveys uncovered a treasure trove of objects in clusters such as 47 Tuc \cite[(Manchester et al. 1991)]{manchester1991} and Ter 5 \cite[(Ransom et al. 2005)]{ransom2005}. The fastest MSP currently known was discovered in Ter 5 with
a spin period of $<$ 1.4 ms \cite[(Hessels et al. 2006)]{hessels2006}.

Other milestones in pulsar discovery include the first eclipsing pulsar \cite[(Fruchter 1998)]{fruchter1988}, that gave us insights into MSP formation and evolution, the first pulsar with a main sequence companion and highly-eccentric orbit \cite[(Johnston et al. 1990)]{johnston1990}, the first planets ever detected beyond our own solar system orbiting PSR B1257+12 \cite[(Wolszczan \& Frail 1992)]{wol1992}, and the pulsars that form the basis of the modern-day pulsar timing arrays, PSRs J0437-4715 \cite[(Johnston et al. 1993)]{johnston1993}, J1713+0747 \cite[(Foster et al. 1993)]{foster1993} and J1909--3744 \cite[(Jacoby et al. 2003)]{jacoby2003}. 

Important pulsars for test of GR include the pulsars with white dwarf companions PSRs J1141--6545 \cite[(Kaspi et al. 2000)]{kaspi2000} and PSR J1738+0333 \cite[(Jacoby et al. 2009)]{jacoby2009}, and the famous double pulsar J0737$-$3039A/B \cite[(Burgay et al. 2003, Lyne et al. 2004)]{burgay2003} (also see Kramer these proceedings). \cite[]{lyne2004}

A revisiting of searching for the single pulses from pulsars resulted in the discovery of the so-called ``Rotating Radio Transients'' or RRATs, and
ultimately the Fast Radio Bursts \cite[(FRBs; Lorimer 2007, Thornton et al. 2003)]{lorimer2007}, 
\cite[]{thornton2013} and the
repeating FRB \cite[(Spitler et al. 2016)]{spitler2016}.

Different sections of the pulsar evolutionary puzzle have ultimately been found, such as the transitional MSP PSR J1023+0038 \cite[(Archibald et al. 2009)]{archibald2009}, and the ``triple system'' PSR J0337+1715 \cite[(Ransom et al. 2014)]{ransom2015}. Among the highest cited pulsar papers are those that give us an insight into the maximum mass of a neutron star such as
PSR J1614--2230 \cite[(Demorest et al. 2010)]{demorest2010} and PSR J0348+0432 \cite[(Antoniadis et al. 2013)]{anton2013}.

\section{The Future of Radio Pulsar Astronomy}

\subsection{Citation and Impact}
Between 1967 and 2017 17,705 papers mentioned ``pulsar'' in the abstract. These papers have a collective 497,442 citations. In 2016 there were over 600 pulsar papers accepted for publication.

A sub-field of astronomy can only thrive if it has impact beyond its own domain. Although pulsar astronomers may be fascinated by certain elements of pulsar phenomenology, the highest impact papers are often those that impact areas of fundamental physics, such as relativity theory, the equation of state of nuclear matter, and whether supernovae are asymmetric?

In a recent review talk, Ken Kellermann listed what he considered radio astronomy's 27 fundamental contributions to science. Only 8 of these had occurred since 1967, despite the field commencing in 1933. His point was that most pivotal discoveries in a new window often occur soon after it is opened. Of the last 8, four involved pulsars - the original discovery, the binary pulsar, the extrasolar planets orbiting PSR B1257+12 and the Fast Radio Bursts. I attribute this to the fact that advances in pulsar science are intimately connected to the available computing power. As Andrew Lyne once explained to me, pulsar surveys are always about signal processing compromises. The phase space we want to search inevitably exceeds our available computational muscle.

Pulsar searches are yet to discover certain jewels such as the greatly anticipated pulsar-black hole binary or a pulsar in a close orbit around the black hole at the centre of our own galaxy. But it is extremely unclear that pulsars can spin significantly faster than what we have already discovered ($P<1.4$ms), or that neutron stars can exceed 2M$_\odot$? Pulsar astronomy's privileged position with regard to tests of strong field gravity is beginning to be eroded by the remarkable discoveries coming from LIGO, including a neutron star binary coalescence. So what does the future hold for this field?

\subsection{The Future of Pulsar Searching}

From 1967 until 2000, the number of pulsar discovered per year was remarkably constant, averaging 22 per year. As the population increases, the number of pulsars that need to be discovered to find something ``interesting'' grows. Fortunately, the pace of pulsar discovery continues to accelerate. From 2000 until 2017 an average of 120 pulsars were discovered each year, and at this meeting a number of interesting pulsars have been revealed, including the slowest pulsar ever detected (Tan these proceedings) and several new relativistic binaries. The pulsar catalogue currently
lists over 2600 pulsars, and when the latest surveys from Green Bank, LOFAR, Parkes and Arecibo are published, this number will exceed 3000.

The MeerKAT, SKA-mid/low, QTT and FAST telescopes all offer hope of accelerating pulsar discovery into the future but these new telescopes are expensive to operate and the science dollar is limited. Telescopes like the SKA will be under immense demand, and it isn't clear that the pulsar science case can demand more that
10-20\% of it.

\subsection{Finding vs Timing}
Although surveys with the SKA-mid and FAST telescopes may yield 1000s of pulsars (e.g. Levin this volume) in surveys that take many months, as telescopes become more sensitive other issues arise. An SKA-mid survey suffers from the fact that most galactic plane survey pointings will contain a pulsar. Historically only a small fraction of survey pointings contain a pulsar, so follow-up timing of all the pulsars takes a few percent of the original survey time. Is a pulsar survey of the plane finds a pulsar in almost every pointing, then follow-up timing becomes comparable to the original survey time, even though most of the pulsars might well be unexceptional. 

The remarkable FAST telescope has exceptional collecting area, but is slow (typically $>$2m) to reposition on the sky. It is clear that a major problem for the future of our field is quickly determining which pulsars are worthy of long timing campaigns.

\section{Predictions}

Since the 2007 Global Financial Crisis in many parts of the Western world science funding is under extreme pressure. Government debt has increased, and science budgets are under the microscope. The cost of construction of radio telescopes that consist of parabolic antennae has not changed that significantly for decades, because the cost of metal and supporting structures is largely invariant with time.

On the other hand, computing power continues its amazing exponential growth. This is moving many telescope designs towards the ``large N-small D'' concept. At this meeting we've seen the new phase space unlocked by LOFAR at low radio frequencies, and the remarkable CHIME telescope (see Ng these proceedings), that takes 1024 inputs to synthesize an enormous number of beams for FRB searches and pulsar timing. CHIME's strength lies in no moving parts that keeps running costs down at the expense of power consumption. It lets the world spin in order to ``slew'' to its targets, and might quickly become the world's leading supplier of pulsar times of arrival in just a year or two.

My own group has been experimenting with the old Mills Cross telescope that possesses 18,000 m$^2$ and synthesizes 512 fan beams for FRB searches and parallel pulsar timing \cite[(Bailes et al. 2017)]{bailes2017}. Telescopes like CHIME and the MOST have very low maintenance costs but can potentially monitor 1000s of pulsars whilst doing other (e.g. FRB) science. Although the pulsar community is a leading SKA advocate, telescopes like CHIME and UTMOST can be constructed for a small fraction of the SKA's annual running cost.

I believe that in the future we are going to see a transition from the very large facilities that move vast amounts of metal to those that possess relatively few moving parts, move electrons to steer, and let the world spin to ``slew'' between sources. Moore's law will allow such instruments to monitor 1000s of pulsars every day. The subsequent data explosion will necessitate artificial intelligence in almost all aspects of our science. From  scheduling through candidate scrutiny, pulsar timing and glitch detection.

\section{Concluding Remarks}

The last 50 years of pulsar science has been amazing and the community a joy to work with. In the next 5 years pulsars can make valuable contributions to areas that range from the equation of state of nuclear matter, gravitational wave sources, the masses, spins and orientations of neutron stars, the dynamics and evolution of pulsars in globular clusters, and the essential correctness of GR to name but a few.

The biggest threats to radio pulsar science are not from ourselves or our scientific peers, but from our dependence on expensive infrastructure. To fund the remarkable tools like the great observatories requires significant investment from our governments. For many years the US has led in the development and provision of large-scale astronomical infrastructure but the polarisation of their electorate into
those that respect science and those that deny it increases
the chance of science funding volatility that is not commensurate with long-term infrastructure planning and maintenance. 

Outside the US there are increasing pressures on governments to fund an aging population with a shrinking tax base and pure research like pulsar astronomy has to compete with research that is seen to have a quicker avenue to ``jobs and growth''. On the other hand, countries that have historically had a small presence in the provision of pulsar observing infrastructure such as Canada, China and South Africa are set to become significant players in the decade ahead.

Finally, a small field like pulsars can do a lot to help itself. Sharing of data, software, students and facilities opens opportunities for the next generation of pulsar astronomers to establish their own groups. The pulsar community came together in Manchester to celebrate a fantastic first 50 years at a great meeting that
was a celebration of our past and a peak at the future. Its greatest asset are the scientists who love the field, and operate with a spirit of mutual respect and cooperation across many countries, forming friendships and striving to outdo each other with great discoveries. It is hard to imagine that the future of this field is anything but bright.

\end{document}